# Alveolar mimics with periodic strain and its effect on the cell layer formation

Milad Radiom[1*], Yong He[2], Juan Peng[2], Armelle Baeza-Squiban[3], Jean-François Berret[1], Yong Chen[2*]

*[1]Université de Paris, CNRS, Matière et systèmes complexes, 75013 Paris, France*
*[2]École Normale Supérieure-PSL Research University, Département de Chimie, Sorbonne Universités-UPMC Univ Paris 06, CNRS UMR 8640, PASTEUR, 24, rue Lhomond, F-75005 Paris, France*
*[3]Unité de Biologie Fonctionnelle et Adaptative, CNRS UMR 8251, Université Paris Diderot Paris-VII, 5 Rue Thomas Mann, F-75205, Paris, France*

**Abstract:** We report on the development of a new model of alveolar air-tissue interface on a chip. The model consists of an array of suspended hexagonal monolayers of gelatin nanofibers supported by microframes and a microfluidic device for the patch integration. The suspended monolayers are deformed to a central displacement of 40 – 80 µm at the air-liquid interface by application of air pressure in the range of 200 – 1000 Pa. With respect to the diameter of the monolayers that is 500 µm, this displacement corresponds to a linear strain of 2 – 10% in agreement with the physiological strain range in the lung alveoli. The culture of A549 cells on the monolayers for an incubation time 1 – 3 days showed viability in the model. We exerted a periodic strain of 5% at a frequency of 0.2 Hz during 1 hour to the cells. We found that the cells were strongly coupled to the nanofibers, but the strain reduced the coupling and induced remodeling of the actin cytoskeleton, which led to a better tissue formation. Our model can serve as a versatile tool in lung investigations such as in inhalation toxicology and therapy.
**Keywords:** Alveolar air-tissue interface; Alveolus mimic; Lung-on-a-chip; Gelatin nanofibers; Physiological strain

## Introduction

Alveolar air sacs are the soft membrane-like tissues at the distal ends of the lungs where gas exchange between atmosphere and blood occurs. The number of the air sacs is in the range of 300 – 400 millions in adult lungs which covers an appreciable surface area of about 70 m$^2$ (Robert, 2000; Weibel, 2015). The air sacs have been shown to have irregular morphologies (Perlman & Bhattacharya, 2007), but on the average to be hexagonal with a diameter of about 200 µm (Fung, 1988). The local epithelium is mainly composed of type I (AETI) and type II (AETII) cells which are respectively squamous and cuboidal in morphology and are respectively responsible for gas exchange and for secretion of lung surfactant fluid to alveolar lumen (Desai, Brownfield, & Krasnow, 2014; Dunsmore & Rannels, 1996; Fehrenbach, 2001; Maina & West, 2005; Weibel, 2015). AETI cells are connected to capillary endothelial cells *via* an extracellular matrix (ECM) which is composed of basement membranes and an interstitial space. The ECM has a thickness of about a few hundred nanometers (Dunsmore & Rannels, 1996; Maina & West, 2005; Townsley, 2012; Weibel, 2015). The composition of the basement membranes are mainly of collagen IV bundles and elastin fibers (Dunsmore & Rannels, 1996; Maina & West, 2005; Townsley, 2012). A thin ECM is required for efficient diffusions of the gas, solutes and proteins across the tissue, and to accommodate tissue expansion during breathing. The secreted lung surfactant fluid facilitates the latter by lowering the air-liquid surface tension to about $10 - 25$ mN/m. From a mechanical point of view, alveoli are the most susceptible tissue in the lungs: they are deformed to a linear strain of $4 - 12\%$ during normal and deep breathing (Fredberg & Kamm, 2006; Guenat & Berthiaume, 2018; Roan & Waters, 2011; Waters, Roan, & Navajas, 2012).

From about three decades ago, several *in vitro* models of alveoli were developed which offer cheap, accessible, easy-to-handle and reliable alternatives to animal models and are used to investigate cell differentiation, surfactant secretion and tissue injury induced by mechanical strain among other investigations (Bilek, Dee, & Gaver, 2003; Chess, Toia, & Finkelstein, 2000; Hermanns et al., 2009; Higuita-Castro, Mihai, Hansford, & Ghadiali, 2014; Jacob & Gaver, 2012; Kamotani et al., 2008; Sanchez-Esteban et al., 2001; Scott, Yang, Stanik, & Anderson, 1993; Trepat et al., 2004; Tschumperlin & Margulies, 1998; Vlahakis, Schroeder, Limper, & Hubmayr, 1999). The use of animal models is unethical and there is a significant social momentum to replace, reduce, and refine them. Over the last decade, microfabrication techniques, such as photolithography, and 3D bioprinting have been used to develop advanced devices for improved *in vitro* modeling by considering topographic cues, shear stress, pressure and deformation, epithelial and endothelial co-culture, amongst others (Campillo et al., 2016; Douville et al., 2011; Higuita-Castro et al., 2017; Huh et al., 2007; Jain et al., 2018; Nalayanda et al., 2009). One model consisted of two overlapping microchannels separated by a microporous polydimethylsiloxane (PDMS) membrane, and used cyclic vacuum to induce uniaxial stretching in the membrane (Huh et al., 2010). This model was used to replicate inflammatory response to tumor necrosis, neutrophil chemotaxis and phagocytosis of bacteria. The same model was later used to replicate pulmonary edema (Huh et al., 2012). Another model incorporated a bio-inspired microdiaphragm in a microfluidic chip in order to induce 3D deformations in a microporous PDMS membrane (A. O. Stucki et al., 2015; J. D. Stucki et al., 2018). These developments are motivated by the need for modeling lung diseases, accelerating the screening of new drugs, and investigating the toxicity of engineered nanomaterials (Park, Georgescu, & Huh, 2019; Tenenbaum-Katan, Artzy-Schnirman, Fishler, Korin, & Sznitman, 2018).

Current *in vitro* models generally use cell culture substrates that are made out of biocompatible polymers or elastomers such as PDMS, polyethylene terephthalate and polyester. Among others, PDMS is more common because it is simple to process, and it has a relatively low elastic modulus leading to its deformability. Nevertheless, PDMS has some shortcomings; *e.g.* it has a high tendency for adsorption of biomarker proteins and drugs which is associated to its hydrophobicity (Boxshall et al., 2006; van Meer et al., 2017), and it does not have the fibrous structure and the biochemical properties of the ECM which are known to direct cell differentiation, migration and other activities (Frantz, Stewart, &

Weaver, 2010; Zhou et al., 2018). To circumvent some of these issues, recently cell culture substrates made out of collagen and elastin were developed and were shown to have a reduced adsorption affinity for biomarkers as compared to PDMS (Zamprogno et al., 2019). Nevertheless, the drop-casted collagen-elastin solutions formed relatively thick layers, 5 – 10 μm, as compared to alveolar ECM (Dunsmore & Rannels, 1996; Maina & West, 2005; Weibel, 2015; Zamprogno et al., 2019).

In the present work, we use an array of suspended monolayers of gelatin nanofibers for replicating the alveolar air-tissue interface. The monolayers are designed to have a hexagonal or honeycomb geometry in order to closely imitate the average shape of the alveolar air sacs (Fung, 1988; Perlman & Bhattacharya, 2007). The use of gelatin is for its biocompatibility and its derivation from collagen which is the most abundant protein in the basement membrane (Dunsmore & Rannels, 1996; Maina & West, 2005; Townsley, 2012). The resulting monolayers have a high porosity which is desirable for a minimal exogeneous material contact with the cells, and a maximal exposure to the culture medium (Liu et al., 2014; Tang, Liu, Li, Yu, Severino, et al., 2016; Tang, Liu, Li, Yu, Wang, et al., 2016; Tang, Ulloa Severino, Iseppon, Torre, & Chen, 2017). The suspended monolayers have a thickness less than 1 μm and a relatively low elastic modulus affording their deformability at the air-liquid interface. We use this property to mimic the breathing dynamics of alveoli by applying air pressure of relatively low values. Human lung epithelial A549 cells originally isolated from a lung tumor (Lieber, Smith, Szakal, Nelson-Rees, & Todaro, 1976) were cultured on the monolayers and exposed to mechanical strain at biologically relevant rates. This cell line is a common model of lung alveolar tissue in *in vitro* investigations (Douville et al., 2011; Higuita-Castro et al., 2014; Huh et al., 2010; Kamotani et al., 2008; Trepat et al., 2004; Vlahakis et al., 1999), and is used here to show the functionality of our alveolar mimics model.

# Materials and methods

*Culture patch.* Fabrication of the culture patch is similar to the previous ones and is schematically shown in **Figure 1(a-j)** (Tang, Liu, Li, Yu, Severino, et al., 2016; Tang, Liu, Li, Yu, Wang, et al., 2016). A silicon wafer is coated with Dry Film resist (DF1100, Engineered Materials Systems, USA) to a thickness of 200 μm using a rolling coater at 100ºC (a). A chromium (Cr) mask with honeycomb pattern (μPG101, Heidelberg Instruments) together with UV exposure are then used to create a honeycomb pattern onto the Dry Film layer (b). On top the exposed Dry Film layer, a new layer of Dry Film is added to a thickness of 100 μm (c). This time, a 2nd Cr mask together with UV exposure are used to pattern an outer ring onto the layer (d). Development in cyclohexanone results in the removal of uncured Dry Film areas (e). The resulting positive mold has a thickness of 200 μm in the central honeycomb area and 300 μm in the outer ring. The resulting mold is initially treated with anti-sticking trimethylchlorosilane by exposure to its vapors for 20 min. Afterwards, PDMS (Sylgrad 184, monomer to curing agent ratio 10:1) is poured on top of the mold (f).

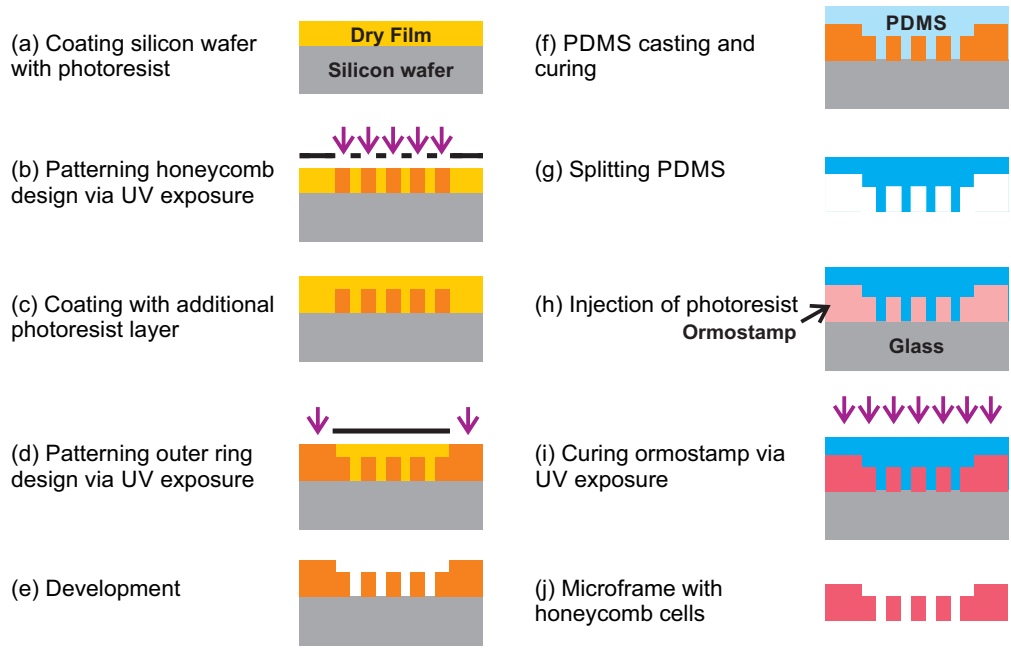

*Figure 1*: Schematic of fabrication steps of honeycomb microframe (a to j) prior to electrospinning of the gelatin nanofibers.

PDMS is cured for 2 h at 70ºC resulting in a PDMS stamp (g). The PDMS stamp is then adhered to glass slide using plasma activation and vacuum. The gap between PDMS stamp and glass is then filled with ormostamp (micro resist technology, Germany) (h) and UV cured for 2 min (i). The result is a microframe with an array of 87 honeycomb cells (j). Each honeycomb has a diameter equal to 500 µm and a pitch distance of 200 µm (**Figure 2(a)**). In the next step, the frame is coated with a thin layer of gold (15 nm) using sputter deposition (Emitech K675X) prior to electrospinning. A gelatin solution containing 10% by weight gelatin (porcine skin, Sigma-Aldrich) in a mixture of acetic acid (840 µl), ethyl acetate (560 µl) and milli-Q water (400 µl) is prepared. During electrospinning, a high voltage (11 kv) is applied to a metal syringe needle while it is placed vertically above the microframe at a distance of 10 cm. A bias is applied to an aluminum foil that is in contact with the microframe from underneath. During electrospinning, the gelatin solution is ejected at a rate of 0.2 ml/h for 4 min. In the next step, electrospinned gelatin nanofibers are dried overnight in a desiccator and then cross-linked in a solution that contains N-(3-dimethylaminopropyl)-N′-ethylcarbodiimide hydrochloride (EDC, Sigma-Aldrich) 0.38 g, and N-hydroxysuccinimide (NHS, Sigma-Aldrich) 0.23 g in absolute ethanol to a total volume of 10 ml. After 4 h of reaction at 4ºC, the culture patches are washed three times in absolute ethanol and let dry overnight in desiccator. The electrospinning and crosslinking steps result in 87 suspended monolayers of gelatin nanofibers on the apical side of the microframe. **Figure 2(b)** shows one suspended monolayer. Each monolayer is hexagonal with a diameter of 500 µm.

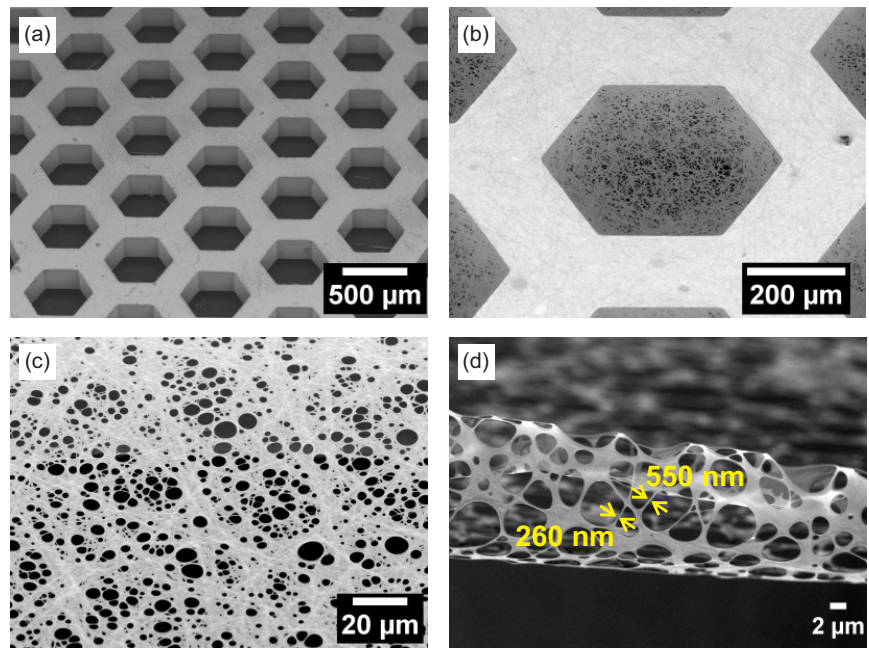

***Figure 2**: (a) Scanning electron microscopy (SEM) image of a culture patch consisting of 87 honeycomb microframes with 500 μm cell size and 200 μm band width. In the picture, 24 microframes are visible. The thickness of the microframes is 200 μm. (b) SEM image of a suspended monolayer of gelatin nanofibers after electrospinning and crosslinking over a honeycomb microframe. (c) SEM image of the microporous structure of a gelatin nanofiber monolayer showing the distribution of pore sizes in the range of 1 – 10 μm. (d) SEM image from a cut through a gelatin nanofiber monolayer showing the thickness of the monolayer in the range of 100 – 500 nm. Except for (c), all images were taken at an angle of 75°.*

*Monolayer deformation.* After fabrication, the patch, with or without A549 cells, is integrated into a microfluidic chip (MesoBioTech) which accommodates air flow from apical channels and medium flow from basal channels. To apply air pressure, the outlet of the air channel is blocked. An air pump is then used to build up air pressure behind an air flow controller (FC-HP1, MesoBioTech), which is then used to regulate and apply air pressure in a sinusoidal wave. The wave function is generated by programming in a dedicated software of the FC-HP1 controller. The maximum pressure is adjusted, and the minimum pressure is set to be the atmospheric pressure. The period of the wave function is set to be 5 s (0.2 Hz) in accordance with the normal breathing rate (Waters et al., 2012). Medium flow is controlled *via* a syringe pump and a flow rate of 0.1 – 1 ml/h is stablished in the basal channels. During operation, the suspended monolayers of gelatin are at the air-liquid interface. *Via* application of air pressure, the monolayers are strained, and their displacements recorded using an inverted optical microscope (Zeiss, Axio Observer.Z1) and a 10× objective lens in phase contrast mode.

*Cell culture.* Adenocarcinoma A549 cells were obtained from the American Type Culture Collection (ATCC, USA). The cells were grown in T25-flasks in Dulbecco's Modified Eagle Medium (DMEM) supplemented with 10% fetal bovine serum (FBS) and 1% penicillin-streptomycin (PS) inside an incubator with humidified atmosphere (37°C, 5% $CO_2$ air). When the cells reached about 80 – 90% confluency they were passaged using TrypLE express enzyme.

Prior to cell seeding, the culture patch was cleaned in 70% ethanol and exposed to UV during 30 min. On the culture patch, the cells were seeded at a density of $6.0 \times 10^5$ cells/ml ($3.0 \times 10^5$ cells/patch). A fresh culture medium was replaced two hours later and then the culture medium was changed every two days. The cells were cultured for 1 to 3 days in submersed state in the incubator (37°C, 5% $CO_2$ air). The culture patch with cells was then mounted in the microfluidic chip and brought to the microscope stage for observations during mechanical strain tests. During these tests, the cells were maintained in the air-liquid interface. After the tests, the culture patch was removed from the chip for the next procedure or analysis, *e.g.* live-dead assay, or put in a petri dish with fresh culture medium for analysis 24 h later. DMEM, FBS, PS and TrypLE were all Gibco brand.

*Live-dead assay*. The culture patch was placed in a petri dish. Cells were incubated with 4 μM Calcein AM (Invitrogen) and 5 μM propidium iodide (Sigma-Aldrich) in Dulbecco's PBS (DPBS) for live and dead cell staining, respectively. After 30 min of incubation (37°C, 5% $CO_2$ air), the cells were rinsed with fresh DPBS two times to remove any residual staining molecules. The fluorescence was measured with an inverted fluorescence microscope (Zeiss, Axio Observer.Z1) using a 10× objective lens. These measurements were repeated on at least two different occasions and on five different cell-cultured monolayers for 1-day and 3-day cultures.

*Immunostaining assay*. The culture patch was placed in a petri dish. Cells were rinsed with DPBS and fixed with 4% paraformaldehyde for 15 min at 4°C. They were then permeabilized using 0.5% Triton X-100 for 10 min, and then saturated with DPBS supplemented with 0.1% Triton X-100 and 3% bovine serum albumin (BSA) for 2 hours at room temperature. Afterwards, the cells were stained with fluorescent antibodies, namely Alexa Fluor 488-conjugated phalloidin (20 μl/ml, Life Technologies, A12379) to label the actin filaments (F-actin), and (4′,6-diamidino-2- phenylindole) (DAPI) to label the nuclei. Samples were observed using a Zeiss confocal microscope (Zeiss Ism 710) and 20× and 40× objectives. These measurements were repeated on at least two different occasions and on three different cell-cultured monolayers for 1-day and 3-day cultures.

*Modeling*. The displacement-pressure response of the gelatin nanofiber monolayers is interpreted in terms of the displacement of a clamped circular plate under uniform pressure. To evaluate the response, the displacement profile of the plate is initially presumed. In the theory that is adapted in this work, the profile has the following expression (Y. Zhang, 2016):

$$w = w_0 \left( 1 - \frac{r^2}{a^2} \right)^2, \tag{1}$$

where $w$ is the axisymmetric displacement of the plate, *i.e.* $w = w(r)$, $w_0$ the displacement at the center of the plate, $r$ the radial distance from the center of the plate, and $a$ its radius.

Solution to the governing equations of plate bending under uniform pressure, gives the following displacement-pressure expression (Y. Zhang, 2016):

$$\frac{Pa^4}{64D} = \left[1 + \frac{3}{4}\left(\frac{a}{t}\right)^2\left(1 - \upsilon^2\right)\frac{\sigma_0}{E}\right]w_0 + \left(0.4118 + 0.25\upsilon - 0.16088\upsilon^2\right)\frac{w_0^3}{t^2}, \tag{2}$$

where $D$ is the bending modulus:

$$D = \frac{Et^3}{12\left(1 - \upsilon^2\right)}, \tag{3}$$

$P$ the pressure, $E$ the elastic modulus, $\upsilon$ the Poisson's ratio, and $t$ the plate thickness. We note that other models are available in the literature and may be used in our data analysis; a list is provided in Ref. (Y. Zhang, 2016). However, the differences between these models estimations are not significant in the calculations of the mechanical properties of PDMS and gelatin nanofiber monolayers that we present in below (Y. Zhang, 2016).

## Results and discussion

*Gelatin nanofiber membrane.* As mentioned before, the fabrication protocol of the culture patch is similar to the previous ones (Tang, Liu, Li, Yu, Severino, et al., 2016; Tang, Liu, Li, Yu, Wang, et al., 2016); however in this work, we changed the material used in making of the honeycomb microframes. In particular, we found that microframes made out of poly(ethylene glycol) diacrylate (PEGDA) swelled in contact with culture medium and changed shape during mechanical strain testing. This situation generally resulted in liquid leaking into the air channels of the microfluidic device. Unlike PEGDA, when ormostamp (micro resist technology, Germany) was used, the microframes were robust and did not change shape. Thereby, our experiments were performed with microframes made out of ormostamp. Ormostamp is a photoresists resin that forms structures with glass-like properties after UV curing. We note that changing the material of the microframe from PEGDA to ormostamp did not affect the process of electrospinning and crosslinking and the general characteristics of the gelatin nanofiber monolayers remained the same (Tang, Liu, Li, Yu, Severino, et al., 2016; Tang, Liu, Li, Yu, Wang, et al., 2016). Moreover, since we found no loss of viability in our cell experiments with ormostamp, this material appears to be compatible with A549 cells as detailed in below.

**Figure 2(a)** shows an array of honeycomb microframes after fabrication *via* lithography. On the basal side of the microframes we electrospinned and then crosslinked a layer of gelatin nanofibers. Before electrospinning, the microframes were coated with a thin layer of gold that set the potential for driving the polymer solution to the microframe during electrospinning. An example of the resulting gelatin nanofiber monolayer is shown in **Figure 2(b)**. This protocol results in an array of suspended monolayers of gelatin nanofibers on the apical side of the microframes. The choice of a honeycomb shape is motivated by the estimated average morphology of the alveoli. Alveolar fields were imaged using confocal microscopy and were shown to consist of irregular morphologies (Perlman & Bhattacharya, 2007); however, the average shape may well be estimated as a hexagon (Fung, 1988). After fabrication, the suspended monolayers have a diameter of 500 μm closely related to the size of alveoli (Robert, 2000; Weibel, 2015). The diameter of an inscribed circle is thereby 433 μm and is used later for the estimations of the elastic moduli and the radial and tangential strains following the theory of equation (2).

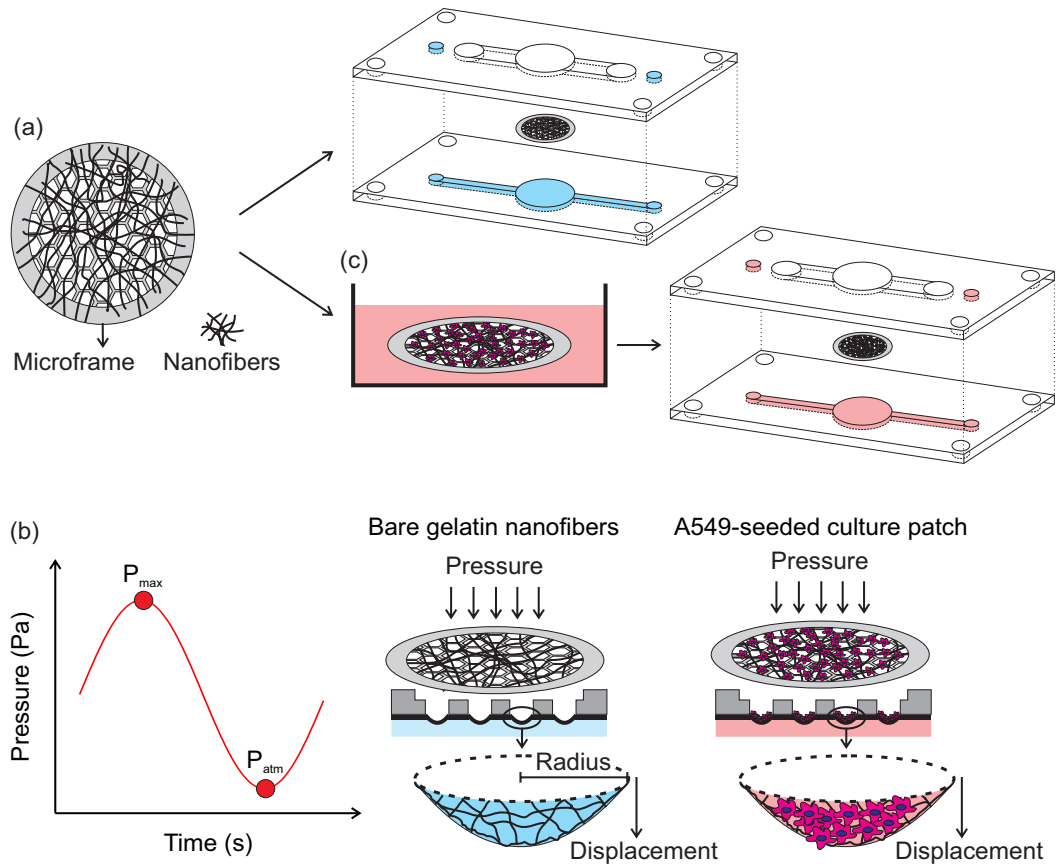

***Figure 3***: *(a) A culture patch consists of honeycomb microframes and suspended monolayers of gelatin nanofibers. The culture patch with or without cells can be reversibly integrated into a microfluidic chip for characterization and evaluation of alveolus mimic functionality. The chip accommodates air pressure from apical channels and liquid flow from basal channels. Water (blue) and culture medium (pink) are used in the experiments with bare patch and cell-cultured patch, respectively. (b) Sinusoidal air pressure wave results in displacement of individual gelatin nanofiber monolayers or cell-cultured monolayers.*

The process of electrospinning and crosslinking results in the formation of a microporous monolayer as shown in **Figure 2(c)**. The porosity is higher in the center of the monolayer whereby the pore sizes are also larger. The porosity decreases gradually with proximity to the microframe walls. The pore sizes were generally in the range of 1 − 10 μm and the porosity was higher than 50%. As a measure of control, we discarded any patch in which the pore size exceeded 20 μm in any of the 87 gelatin nanofiber monolayers in one culture patch. We cut a few patches in order to measure the thickness of the monolayer. Interestingly we found that after cutting the monolayer folds on itself. This behavior shows a pre-stress state in the gelatin nanofibers induced during crosslinking. After being cut, the stress is released, and the monolayer folds on itself. To measure the thickness, we set the angle of scanning electron microscope (SEM) camera to 75° and measured thickness values in the range of 100 – 500 nm as shown in **Figure 2(d)**. The thickness of the monolayer is thereby in the range of the thickness of alveolar ECM. Moreover, the monolayer microporous and fibrous structure resembles ECM structure (Dunsmore & Rannels, 1996; Maina & West, 2005; Townsley, 2012; Weibel, 2015). Since gelatin is derived from collagen, the resulting substrate has in addition an ECM-like chemical composition (Dunsmore & Rannels, 1996; Maina & West, 2005; Townsley, 2012; Weibel, 2015). These characteristics offer advantages to PDMS-based models. Additional images of the gelatin monolayers, distribution of pore sizes, and monolayer thickness are provided in Supporting Information Part 1.

*Biomimetic strain.* A culture patch was integrated in a microfluidic chip with basolateral and apical channels which were used respectively for medium flow and air pressure. In the experiments with bare culture patch, *i.e.* with no cell seeding, water was used as the medium. After filling the basolateral channels with water, the monolayers were at the air-liquid interface. Application of air pressure in this case resulted in the deformation of the monolayers. This setup is shown schematically in **Figure 3**. During these tests, 5 monolayers were randomly selected and for each monolayer the displacement of the center $w_0$ was recorded. After removal of pressure, we ascertained that the monolayers returned to their initial undeformed state.

**Figure 4** shows the displacement of the gelatin nanofiber monolayers as a function of the maximum of the applied air pressure. Each data point is an average of central displacement from 5 different monolayers and the error bars are the standard deviations. The differences in the displacement responses of individual monolayers at each pressure, *i.e.* the extent of standard deviations, may be associated with differences in the density of the nanofibers in the monolayers as well as thickness. However, these variations may be estimated to be minute considering that electrospinning is a random deposition process and may be improved if the fibers were deposited uniformly for example by rotating the microframe during electrospinning. Nevertheless, from the standard deviation values we find that the monolayers vary by only about 10% or less from the average response which we consider to be insignificant. The resolution in the pressure measurements was estimated to be about 100 Pa; this value is added as an error bar to the data points. We find that it is generally possible to exert pressures up to about 1000 Pa and induce displacements as large as 100 μm.

The displacement of the monolayer at the air-liquid interface results from an interplay between the applied air pressure, the elasticity of the monolayer, and the liquid surface tension. In our measurements, we obtain an overall displacement which is precisely a cooperative displacement of the monolayer and the air-liquid interface. The displacement of the air-liquid interface may be calculated from Laplace's equation (Israelachvili, 2011):

$$P = \gamma \left( \frac{1}{R_1} + \frac{1}{R_2} \right), \tag{4}$$

where $\gamma$ is the liquid surface tension equal to 72.8 mN/m for water and $R_1$ and $R_2$ are the two orthogonal or principal radii of curvature. In our geometry, $R_1 = R_2$. Using chord theorem, equation (4) simplifies to:

$$P = \frac{4\gamma w_0}{a^2 + w_0^2}.$$ (5)

In **Figure 4** we show that the overall displacement is smaller than the prediction of Laplace's equation. This difference reveals the effect of gelatin nanofibers in stiffening the interface. We performed control experiments with honeycomb microframes that did not have the gelatin monolayers and observed that beyond a pressure of about 400 Pa the air broke into the liquid channel. This observation shows that the air-liquid interface becomes unstable at pressures higher than 400 Pa. Addition of gelatin nanofiber monolayers to the air-liquid interface stabilizes the interface whereby pressures more than 1000 Pa may be applied. Although the values of the interface displacement become smaller, it is explained below that the resulting strains in the monolayers are in the range of the physiological alveolar strain.

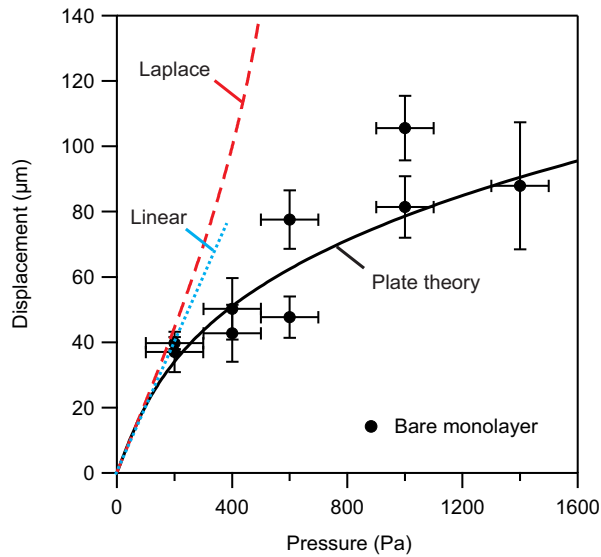

***Figure 4**: Displacement-pressure response of gelatin nanofiber monolayer together with fit to plate theory equation (2). The best fit values are the elastic modulus $E = 4.0 \pm 1.0$ MPa and the residual stress $\sigma_0 = 0.2 \pm 0.1$ MPa, while the Poisson's ratio is set to 0.5. The linear response approximation for low values of applied pressure is also depicted (blue dotted line). Displacement-pressure response of air-liquid interface is calculated from Laplace's equation (5) and is depicted for comparison (red broken line). Error bars on y-axis are standard deviations of recorded displacements from 5 different suspended gelatin monolayers. Error bars on x-axis are error of pressure measurement estimated to be about 100 Pa.*

**Figure 4** further shows that the displacement of the monolayer increases with pressure but deviates from a linear dependence. In the theory of plate deformation, this departure is associated with transition from bending-dominant displacement to stretching-dominant displacement (Wan, Guo, & Dillard, 2003; Wan & Lim, 1998; Y. Zhang, 2016), and is found to occur early in our applied pressure range. The displacement values as a function of pressure were fit to equation (2) as shown in the figure. The fit parameters are the elastic modulus and the residual surface stress for which we obtained respectively $E = 4.0 \pm 1.0$ MPa and $\sigma_0 = 0.2 \pm 0.1$ MPa. We did not fit for the Poisson's ratio and instead assumed $\upsilon = 0.5$. The elastic moduli of wet gelatin scaffolds and membranes have been measured in the past using methods such as stress-strain mechanical testing and found to be in the range of $1 - 4$ MPa in agreement with our measured value (Higuita-Castro et al., 2017; Kim et al., 2009; S. Zhang et al., 2009). We note that although the measured value for the elastic modulus of gelatin monolayer shows that this construct is compliant (*e.g.* as compared with PDMS), this value is still very high as compared with the elastic modulus of lung tissue which has been measured to be in the range of a few kPa (Polio et al., 2018; Roan & Waters, 2011).

An important conclusion from this part is that the suspended gelatin monolayers at the air-liquid interface can act like membranes with a well-defined displacement-pressure response. For example, similar experiments were performed with PDMS membrane in the same microfluidic chip. The PDMS membranes had a diameter equal to 9 mm and thicknesses in the range of $50 - 400$ µm. A sinusoidal air pressure wave was applied to the apical side of the membranes while water flew through the basal channels. Air pressure resulted in the displacement of the PDMS membrane, for which we recorded the value at the center. Since PDMS is transparent, we used a marker to color the center of the membrane in order to ease its visualization in the microscope. We found that displacement increases with pressure but deviates from a linear dependence which is again associated with stretching dominant displacement (Wan et al., 2003; Wan & Lim, 1998; Y. Zhang, 2016). Fits to the displacement values using equation (2) resulted in elastic moduli in the range of $3 - 7$ MPa in agreement with the reported values in literature (Qian et al., 2016; Thangawng, Ruoff, Swartz, & Glucksberg, 2007). Experiments with PDMS membranes are presented in Supporting Information, Part 2.

The evolution of displacement with pressure that we obtained with gelatin monolayers at the air-liquid interface as shown in **Figure 4** has a similar qualitative behavior to PDMS membranes (Supporting Information Figure SI 3). This similarity corroborates that the mechanical response of the microporous monolayer is akin to a membrane making the monolayers structurally, *i.e.* ECM-like, and mechanically, *i.e.* membrane-like, relevant for alveolus mimic investigations as shown below.

We now come back to the discussion of mechanical strain as a result of the monolayer displacement. In Supporting Information SI 3 we provide a full mechanical characterization based on the theory of equation (2). The equations for radial and tangential strains are also provided. From the calculations, one obtains that the maximum of the radial and tangential strains are $e_{\text{r,max}} = 0.7 \times (w_0/a)^2$ and $e_{\text{t,max}} = 0.6 \times (w_0/a)^2$, respectively. However, these maxima do not occur in the same radial position. In particular, the radial strain is maximum at a distance $r = 0.6a$ closer to the periphery, while the tangential strain is maximum in the center. Applying these relations, we find a maximum strain value in the range of $2 - 10\%$ from the displacement-pressure response of the monolayers shown in **Figure 4** which is in good agreement with the physiological range of alveolar strain, 4 – 12% (Fredberg & Kamm, 2006; Guenat & Berthiaume, 2018; Roan & Waters, 2011; Waters et al., 2012).

*Alveolus mimic model.* The gelatin nanofiber monolayers were then used as substrates for the culture of A549 cells. This cell line is commonly used in *in vitro* investigations of lung alveoli as related with technical developments of microfluidic devices, as well as modelling cell responses (Douville et al., 2011; Higuita-Castro et al., 2014; Huh et al., 2010; Kamotani et al., 2008; Trepat et al., 2004; Vlahakis et al., 1999). The cells were seeded at a density of $3.0 \times 10^5$ cells/patch and were found to start adhering to gelatin nanofibers within an hour after seeding. A culture of one day generally resulted in a cell layer with about 50% confluency and a distribution of both extended and round morphologies. After three days most cells were adhered well to the gelatin nanofibers and the confluency was more than 90%. Comparing to the growth on a petri dish, we find a similar growth rate with a duplication time of about 1 day in good agreement with ATCC. Examples of cell culture of gelatin monolayers are presented in Supporting Information SI 4.

Thereafter, cell-cultured patches were integrated into the microfluidic chip for the application of air pressure and mechanical strain as depicted schematically in **Figure 3**. Air pressure wave was applied from the apical side and a flow of culture medium was stablished in the basal channels. Recording cell layer displacement was similar to the bare gelatin monolayers however the presence of the cell layer made the visualizations easier.

**Figure 5(a)** shows a series of light microscopy images where the displacements of a cell-cultured gelatin monolayer are shown. One observes that from the initial relaxed position (i), where the central displacement $w_0 = 0$ µm at the atmospheric pressure, the entire surface of the cell-cultured monolayer deforms by increasing air pressure. At a pressure of about half of the maximum, the center of the monolayer is displaced by 35 µm (position ii) while at the maximum pressure of 400 Pa the center of the monolayer is at its full-depth displacement which is measured to be 70 µm (position iii). In Supporting Information SI 5, we show that the variation in the transmitted light intensity between the center of the monolayer and its corners correlates linearly with the central displacement. We use the linear relation to calibrate the displacement of the entire surface area of the monolayer. **Figure 5(b)** shows the displacement of the monolayer along its diameter at similar positions i to iii.

The displacement profiles shown in **Figure 5(b)** may then be interpreted in terms of the radial displacement profile of equation (1). As shown in the figure, there is a good agreement between the experimental profiles and the model validating the use of equation (2) in fitting our data. More importantly the theory may be used to estimate the extent of radial and tangential strains during the experiments which we will discuss in below.

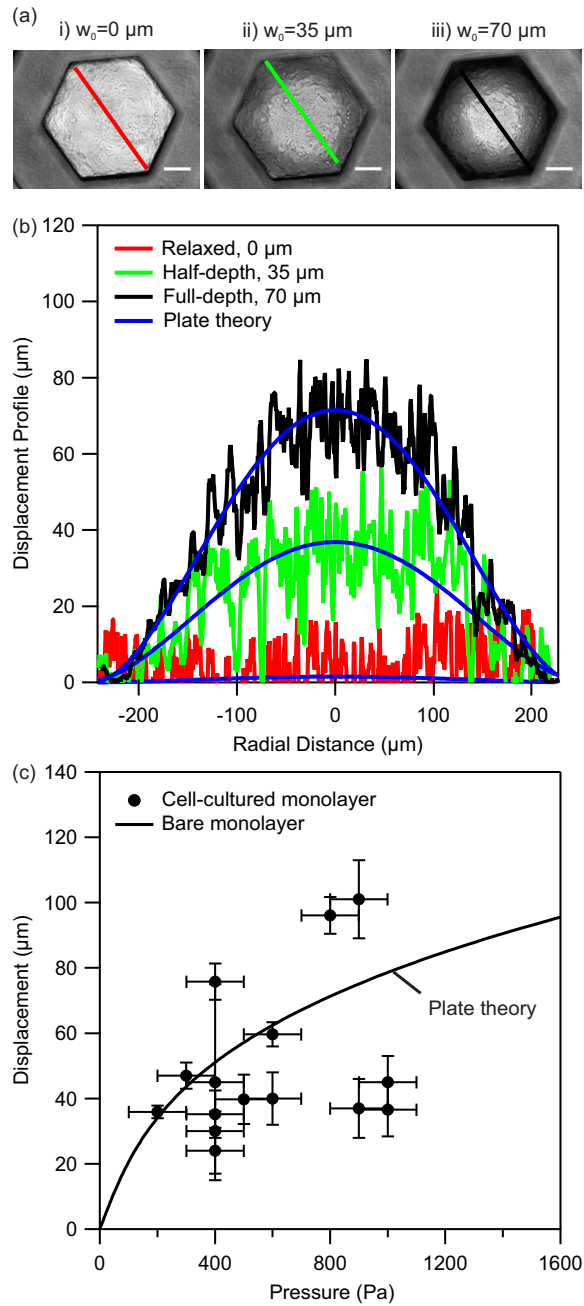

**Figure 5**: _(a) Light microscopy images of a cell-cultured gelatin monolayer at different displacement levels. Position 'i' is the initial relaxed state at the atmospheric pressure, position 'ii' the half-depth displacement equal to 35 µm, and position 'iii' the full-depth displacement equal to 70 µm at the maximum of the applied air pressure equal to 400 Pa in this experiment. (b) The displacement profile of the cell-cultured gelatin monolayer obtained from the variations in the light intensity from the center to the edge of the monolayer after calibration (SI, Part 5). The displacement profile shows a good agreement with the theoretical profile of equation (1). (c) Displacement-pressure response of several cell-cultured gelatin monolayers. Solid line is the reproduced displacement-pressure response of bare gelatin monolayer obtained from the fit to data in **Error! Reference source not found.**. Error bars on y-axis are standard deviation of recorded displacements from 5 different cell-cultured monolayers. Error bars on x-axis are error of pressure measurement estimated to be about 100 Pa. In total seven cell-cultured patches were investigated._

Several cell-cultured patches were then investigated for their displacement-pressure responses. For each culture patch, we measured the displacements of 5 different cell-cultured monolayers at different locations on the culture patch and report an average value with a standard deviation in **Figure 5(c)**. The extent of standard deviation has the same argument as mentioned earlier in the case of bare gelatin monolayers. However, with cells, there is an additional complexity due to the use of culture medium with low surface tension which facilitates occasional formation of small bubbles in the channels. These bubbles may interfere with the displacement of the monolayer and increase the extend of variations in the measured displacements. Nevertheless, we find that for a pressure range of $400 - 1000$ Pa, the cell layer is displaced in the range of $20 - 100$ μm. In **Figure 5(c)** we included the fitted curve to the displacement-pressure response of bare monolayers from **Figure 4**. Comparison of the experimental displacement-pressure response of cell-cultured monolayers with the response of monolayers without cells shows similarity in the range of displacements as a function of pressure between the two cases. This observation shows that the added cell layer did not significantly alter the elasticity of the monolayer. We attribute this observation to a much higher compliance (*i.e.* less stiff character) of the cell layer as compared with the gelatin monolayer (He, Chen, Sun, & Zheng, 2012). In particular, the elastic modulus of single cells and cell monolayers have been measured to be in the range of a few kPa using atomic force microscopy nanoindentation technique and stress-strain mechanical tests (Brückner & Janshoff, 2015; Harris et al., 2012; Roan, Wilhelm, & Waters, 2015). The stiffness of the cell monolayer is thereby much smaller than the stiffness of the gelatin monolayer for which we measured a value of elasticity equal to 4 MPa. Thereby, the combination of the two layers has an elastic modulus that is similar to the elastic modulus of gelatin monolayer. This finding is similar to those reported in the past with silicon membranes and culture of AETII rat cells (Tschumperlin & Margulies, 1998). This characteristic of the gelatin monolayer is akin to ECM that is also responsible for the mechanical integrity and stability of the alveolar tissue during breathing extensions (Dunsmore & Rannels, 1996; Maina & West, 2005; Waters et al., 2012; Weibel, 2015).

We now come back to the discussion of the effect of mechanical strain on the cells as a result of the monolayer displacement. The equations for radial and tangential strains were shown earlier. Applying these relations, we find a maximum strain value in the range of $1 - 11\%$ from the displacement-pressure response of cell-cultured monolayers shown in **Figure 5**. This range is again in good agreement with the physiological range of alveolar strain, $4 - 12\%$ (Fredberg & Kamm, 2006; Guenat & Berthiaume, 2018; Roan & Waters, 2011; Waters et al., 2012). We discuss cell viability and morphology in the following sections.

*Cell viability.* Several culture patches were tested for evaluating cell viability after the mechanical strain tests. These patches were deformed to a strain of about 5% at a frequency of 0.2 Hz for a duration of 1 h. Immediately after the tests (denoted by 0 h), the patches were stained with markers of live and dead cells, calcein AM and propidium iodide respectively, and observed with fluorescence microscopy. **Figure 6(a)** shows examples of these measurements. The green emission originates from cellular enzymatic activity with calcein and is a signature of live cells. We find a significant number of live cells on the gelatin monolayers after the strain tests. The effect of the strain is still not significant after 24 h, where again a significant green emission is collected from the same patch. Limited number of red emissions is associated with propidium iodide diffusing inside a permeable membrane and signifies a low number of dead cells. We quantified the total number of cells by summing up the total number of live and dead cells and expressed the cell viability in terms of the percentage of the live cells. **Figure 6(b)** and **Figure 6(c)** show that no significant reduction in cell viability was found upon application of mechanical strain and in particular in comparison with parallel measurements with cells that did not experience any strain. The results persist in cultures of 1 day and 3 days. Slight increase in the intensity of the reflected green light at the periphery of the monolayer in **Figure 6(a)** is not related with the mechanical strain and is seen with static patches as well. A higher number of cells in the periphery may be related with the smaller pore sizes in this part of the monolayer which provides a larger number of adhesion sites for the cells. This effect may result in cell migration or higher proliferation in the periphery.

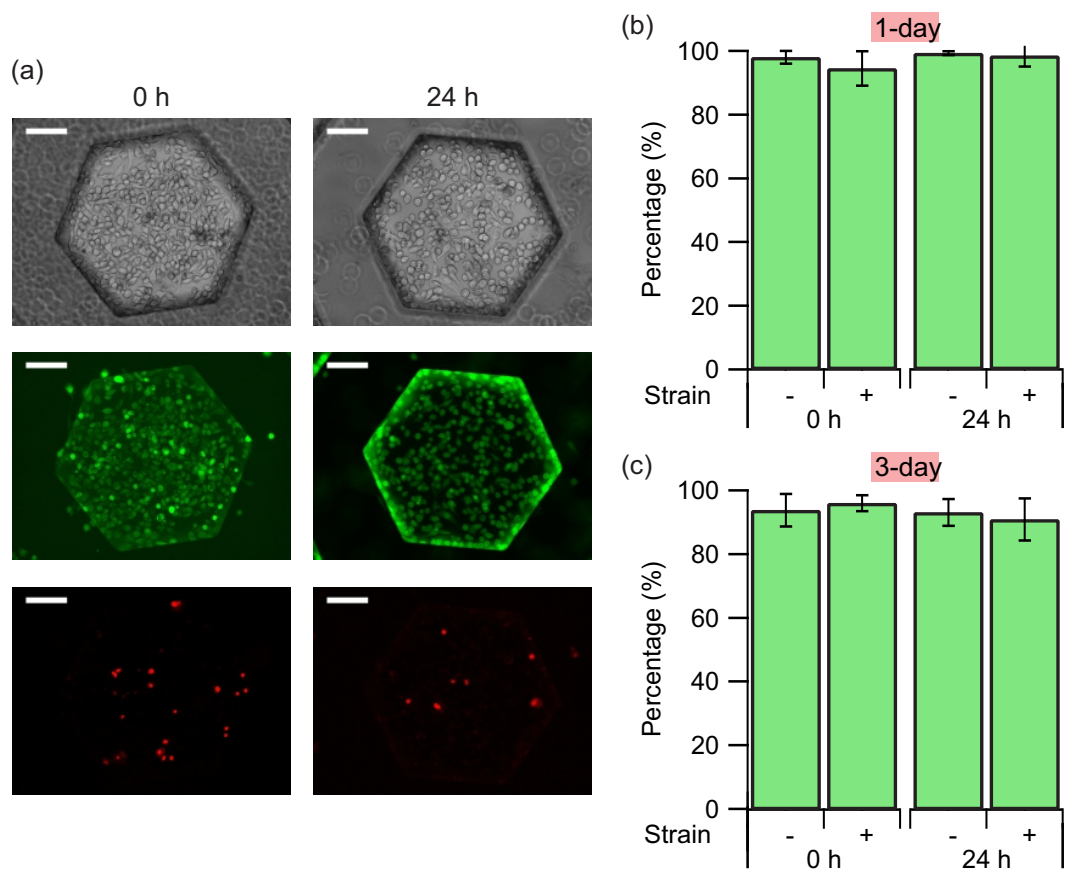

***Figure 6***: *(a) Cell viability assay using calcein to mark live cells (green) and propidium iodide to mark dead cells (red) immediately after the strain test denoted by 0 h and after incubation for 24 h. The top images are phase contrast images of the cells followed by green and red fluorescence emission images showing the distribution of the live and the dead cells respectively. Scale bar, 100 µm. (b, c) Deduced percentages of viable cells at day 1 (b) and day 3 (c), with and without application of a periodic strain of 5% at a frequency of 0.2 Hz for a duration of 1 h. Values are*

*expressed as mean ± standard error of mean. These measurements were repeated on at least two different occasions and on five different cell-cultured monolayers for 1-day and 3-day cultures.*

The results of **Figure 6** are in agreement with a previous work which showed that mechanical strain had no effect on the proliferation and viability of A549 cells at higher strain values than used here (Vlahakis et al., 1999). It was otherwise shown that mechanical strain may increase the proliferation of the adenocarcinoma alveolar H441 cells (Chess et al., 2000), and the primary human alveolar epithelial cells (A. O. Stucki et al., 2015). Slightly increased viability after mechanical strain was also observed with primary human bronchial cells (J. D. Stucki et al., 2018). Measurements with AETII cells from rat models showed that seeding density and incubation time may have an effect on cell death during mechanical strain tests (Tschumperlin & Margulies, 1998). In particular, cell death was found to be the least when the seeding density was the highest, an effect which was associated with increased cell-cell contacts (Tschumperlin & Margulies, 1998). Increasing strain level to higher than the physiological range is otherwise shown to result in tissue injury and cell death (Hammerschmidt, Kuhn, Grasenack, Gessner, & Wirtz, 2004; Tschumperlin, Oswari, & Margulies, 2000). Substantial injuries are reported in models of air-way re-opening in which shear stresses from moving droplets are applied to the cells (Bilek et al., 2003; Douville et al., 2011; Higuita-Castro et al., 2014; Higuita-Castro et al., 2017). From our results, we conclude that a physiological strain of 5% has no effect on the viability of A549 cells. This result further shows the applicability of our model in replicating the environment of alveolar cells.

*Cell morphology.* Cell behaviors are sensitive to stiffness, morphology and biochemical composition of the substrate. On a suspended monolayer of gelatin nanofibers, cells tend to explore adhesive contacts, resulting in cell membrane deformation and an increase in contact area with the nanofibers. If the cell density on the monolayer is high enough, cell-cell contacts become important at the expense of adhesion with the nanofibers (Dahlin, Kasper, & Mikos, 2011; Wu et al., 2011). When a periodic deformation is applied, the cell cytoskeleton is reorganized due to mechanotransduction. We investigated the effect of periodic strain on actin cytoskeleton rearrangement which is the most studied cytoskeleton element. We examined these effects by analyzing immunofluorescence images of A549 cells on the culture patch after incubation for 1 and 3 days and applying a periodic strain of about 5% at frequency of 0.2 Hz. The strain level was chosen to be in the range of alveolar strain during normal breathing and the applied frequency to match the normal breathing rate (Fredberg & Kamm, 2006; Guenat & Berthiaume, 2018; Roan & Waters, 2011; Waters et al., 2012). Pressure application continued for a period of 1 h which has been shown to prevent tissue damage at similar strain levels (Tschumperlin & Margulies, 1998; Tschumperlin et al., 2000). Similar test conditions have been used in modeling nanoparticle interactions with lung alveoli, mechanical ventilation-induced alveolar tissue injury, as well as obstructive sleep apnea (Campillo et al., 2016; Davidovich et al., 2013; Huh et al., 2010). After the strain tests, the cells were immediately fixed and stained with phalloidin and DAPI to mark actin filaments in green and nuclei in blue, respectively.

**Figure 7** shows the effect of strain on the cell layer formation. The immunostaining images were acquired with cells on gelatin monolayers after 3 days of incubation without application of mechanical strain as shown in **Figure 7(a)**. In parallel, images were obtained will cells after 3 days of incubation and after application of strain as shown in **Figure 7(b)**. A few of image sections were selected for the close comparison of in-plan and out-plan cell distributions. Images marked Z = 0 indicate top layers (out-plan), the ones labeled Z = +3 μm or +6 μm show the main layers (in-plan), and the other sections labeled Z = +10 μm represent bottom layers corresponding to cell-nanofiber contact (out-plan). The parallel examination shows that without the strain step, the cell distribution was not homogenous on the gelatin monolayer and some cells went out of the main plan. On the contrary, the strain step resulted in a monolayer of cells in the whole honeycomb compartment, indicating that the periodic strain may release the cells from their original anchor points with the nanofibers resulting in a more homogenous redistribution which indicates the formation of a tighter cell layer.

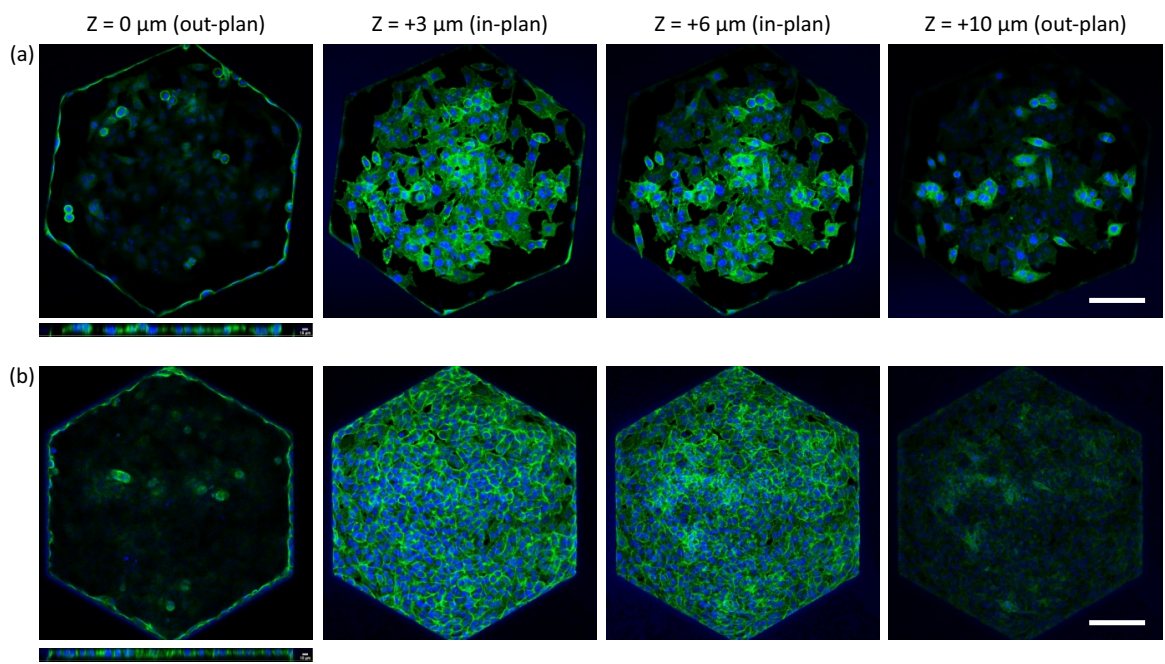

*Figure 7: Selected sections of immunostaining confocal images of cells on culture patch after incubation for 4 days: (a) without mechanical strain, and (b) with a periodic strain (level 5% at a frequency of 0.2 Hz for a duration of 1 h) at day 3. F-actin was stained with Alexa Fluor 488-conjugated Phalloidin (green) and nuclei with DAPI (blue). Scale bar, 100 μm.*

**Figure 8(a)** shows examples of the confocal images of the cells found at the center and the peripheral parts of the monolayers. The images were taken at different height levels, and in this figure, 'bottom' corresponds to cell-nanofiber contact and 'maximum' to maximum intensity projection of pixel value. Selected areas from 'bottom' images are enlarged and shown to the left of these images. The enlarged areas are marked with dashed line squares. In Supporting Information SI 6 additional images are shown at the level of the nuclei which is about 5 µm above the cell-nanofiber contact and denoted by 'middle', and at a location that is 12 µm above the contact and denoted by 'top'.

Cells that did not experience any strain after 1 day of culture displayed a number of green fluorescence accumulations in the form of spots in all cell areas. These spots are visible in cell-nanofiber contact level images and are marked with yellow arrows in **Figure 8(a)** Formation of these accumulations indicate a non-homogeneous distribution of actin filaments (F-actin) at the cell-nanofiber contact level. Due also to a strong cell-nanofiber contact, cells generally display a non-rounded morphology which is more clearly visible in the nuclei level images shown in Figure SI 6. Cells of 1-day culture that experienced strain show a reduced number of F-actin spots particularly for the cells located in the central parts of the monolayer where the strain is the highest.

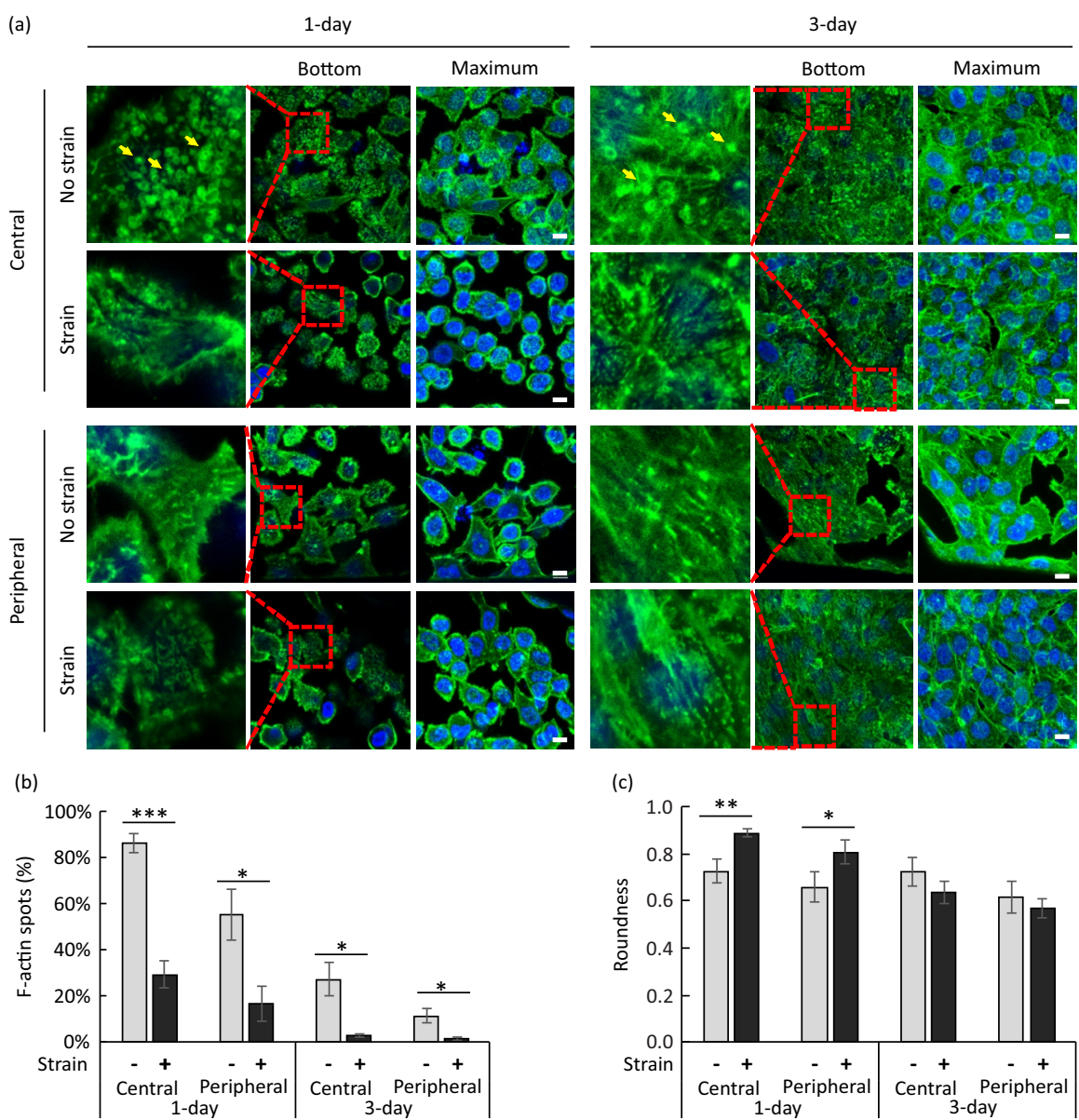

*Figure 8*: (a) Confocal images of immunostained cells on culture patches after 1 day and 3 days of culture in static condition and with a periodic strain of 5% at a frequency of 0.2 Hz for a duration of *l h. Images were taken from the cells in the central and the peripheral parts of the gelatin monolayers. Single confocal sections from the cell-nanofiber contact level ('bottom') and maximum intensity projections of z-stacks ('maximum') are shown. Dashed line squares on 'bottom' images show selected areas which are enlarged and shown to the left. Yellow arrows point at green fluorescence accumulations which are observable in the enlarged areas. F-actin was stained with Alexa Fluor 488-conjugated Phalloidin (green) and nuclei with DAPI (blue). Scale bars, 10 μm. (b) Deduced percentage of cells with clear F-actin accumulations in the cell areas at the cell-nanofiber contact level at day 1 and day 3 in static condition and with strain (5%, 0.2 Hz, 1 h). (c) Deduced cell roundness at day 1 and day 3 in static condition and with strain (5%, 0.2 Hz, 1 h). The measurements were repeated on at least two different occasions and on three different cell-cultured monolayers for 1-day and 3-day cultures. Values are expressed as mean ± standard error of mean. \*P < 0.05, \*\*P < 0.001, \*\*\*P < 0.0001 (Student's t-test).*

**Figure 8(b)** shows the effect of strain on the percentage of cells with clear F-actin accumulations in the cell areas inspected at the level of cell-nanofiber contact. Application of strain is found to reduce the number of clear F-actin accumulations in cells of 1-day as well as 3-day cultures. The effect is found to be significant in both the central and peripheral parts of the gelatin monolayer, although more cells with clear F-actin accumulations are generally found in the central part than the peripheral part. The latter effect is associated with denser fibrous parts in the periphery. Clear F-actin spots became also less visible with incubation time which is associated with effects from increased cell-cell contacts with a longer incubation.

**Figure 8(c)** shows the effect of strain on the cell morphology. We express cell morphology in terms of a roundness parameter, which is defined by $4A/\pi d^2$, where $A$ is the cell area and $d$ the length of the major axis. We find that with cell cultures of 1 day, the cell morphology is largely influenced by the nanofibers' network due to the effect of contact guidance. For example, a periodic deformation of the network at a physiological strain level of 5% is sufficient for remodeling the cytoskeleton and to make the cells more rounded. With high cell number on the monolayer, the cell-cell interactions become important and reduce the influence of the underneath mobile network on changing the cell morphology. Although strong cell-nanofiber adhesion prohibits cell movements, periodic deformation helps the cells to move and to form a homogeneous epithelial layer in the central and peripheral parts of the monolayer. The result is that the cells appear to be less rounded after the strain step in 3-day cultures which is evident from a decrease in roundness parameter of the cells. These results corroborate previous observations of good adhesion between gelatin nanofibers and other cells such as human pluripotent stem cells, motor neuron cells, primary hippocampal neurons and cardiomyocytes (Liu et al., 2014; Tang, Liu, Li, Yu, Severino, et al., 2016; Tang, Liu, Li, Yu, Wang, et al., 2016; Tang et al., 2017), and extends the applicability of gelatin nanofiber monolayers to alveolar cells.

## Conclusion

We have demonstrated that suspended monolayers of gelatin nanofibers could be used as membrane-like substrates for alveolar cell culture and periodic deformation to mimic the alveolar air-tissue interface. Comparing to the more conventional approaches using PDMS membrane, this technique is advantageous: gelatin is derived from collagen, the monolayer has a microporous structure providing the possibility of the air-liquid interface culture and minimal material contact with the cells, the monolayer has a relatively low elastic modulus resulting in its strain modulation in a physiological range. A549 culture on the monolayers showed viability after periodic strain of 5% at 0.2 Hz during 1 h for cultures of 1 day and 3 days. Our results showed that the strain can efficiently remodel the actin cytoskeleton of the cells in the cell-nanofiber contact positions by reducing the number of the anchoring points leading to a better tissue formation. The herein developed methods and tools are applicable to diverse lung studies *e.g.* inhalation toxicology and therapy.

**Acknowledgements.** Maria Sarkis, Oliver Brookes and Elrade Rofaani are thanked for fruitful discussions. Ayako Yamada is thanked for help during fabrication of honeycomb microframe. Agence Nationale de la Recherche (ANR) is gratefully acknowledged for their financial support of this work through ANR-17-CE09-0017 (AlveolusMimics). The authors declare no conflict of interest.